  \documentclass[final,3p,times,twocolumn]{elsarticle}%
\usepackage{amsmath,amssymb,exscale}
\usepackage{color}
\usepackage{hyperref}%, pdfsync, epstopdf, enumerate
\hypersetup{colorlinks,bookmarksopen,bookmarksnumbered,citecolor=rossos,
linkcolor=black,pdfstartview=FitH,urlcolor=blus}
\usepackage{slashed}

\definecolor{blus}{cmyk}{1,1,0,0.6}
\definecolor{verdes}{cmyk}{0.99,0,0.59,0.82}
\definecolor{rossos}{cmyk}{0,1,1,0.55}
\definecolor{greeny}{cmyk}{0.99,0,0.59,0.98}

\def\be{\begin{equation}}
\def\ee{\end{equation}}
\def\bea{\begin{eqnarray}}
\def\eea{\end{eqnarray}}

\def\dsl{\hbox{\hbox{${\partial}$}}\kern-1.7mm{\hbox{${/}$}}}
\def\Dsl{\hbox{\hbox{${D}$}}\kern-1.9mm{\hbox{${/}$}}}
\definecolor{red}{rgb}{1,0,0}
\journal{the arXiv}

\begin{document}

\begin{frontmatter}



\title{A Simple Motivated Completion of the Standard Model below the Planck Scale: \\ Axions and Right-Handed Neutrinos}
%title for the letter ""

%\author{Alberto Salvio}

\author{\large Alberto Salvio}

\address{\normalsize Departamento de F\'isica Te\'orica, Universidad Aut\'onoma de Madrid\\ and Instituto de F\'isica Te\'orica IFT-UAM/CSIC,  Madrid, Spain. \\ {\it {\small Report number: IFT-UAM/CSIC-15-003}} }

\begin{abstract}
We study a simple Standard Model (SM) extension, which includes three families of right-handed neutrinos with generic non-trivial flavor structure and an economic implementation of the invisible axion idea. We find that in some regions of the parameter space this model accounts for all experimentally confirmed pieces of  evidence for physics beyond the SM: it explains neutrino masses (via the type-I see-saw mechanism), dark matter, baryon asymmetry (through leptogenesis), solve the strong CP problem and has a stable electroweak vacuum. The last property may allow us to identify the Higgs field with the inflaton.\end{abstract}


\end{frontmatter}

% \linenumbers

% main text

\section{Introduction}

Although no unambiguous signal of physics beyond the SM (BSM) has appeared so far at the LHC, there is no doubt that the SM has to be extended. Neutrino oscillations, which lead to the existence of small (left-handed) neutrino masses, and the observational evidence for dark matter (DM) is enough to state that the SM is incomplete. 

Other unsatisfactory features of the SM are an insufficient  baryon asymmetry of the universe, the strong CP, gauge hierarchy and cosmological constant problems. 

Moreover, precision calculations \cite{Degrassi:2012ry,Buttazzo:2013uya} indicate that   the SM potential develops  an instability at a scale  of the order of $10^{10}$ GeV, for central measured values of the SM parameters. This is not particularly worrisome {\it per se} because the probability of tunneling to the absolute minimum, where life is impossible, is spectacularly small \cite{Buttazzo:2013uya}. However, it may lead to some issues during  the exponential expansion of the early universe (inflation) 
\cite{Hook:2014uia,Goswami:2014hoa,Herranen:2014cua}.  
  Moreover, the (absolute) stability up to the Planck scale  $M_{\rm Pl}$ may lead to the possibility of Higgs inflation  \cite{Bezrukov:2007ep,Bezrukov:2008ej, Bezrukov:2012sa,Salvio:2013rja}, linking particle physics and cosmology: this is interesting because it provides us with relations between particle physics and cosmological observables.  The presence of such an instability in the SM is not firmly confirmed because of non-negligible uncertainties on the top mass and the QCD gauge coupling; but, if confirmed, it would  suggests that  right-handed neutrinos (at scales suitable for the see-saw mechanism and thermal leptogenesis) and the physics of the QCD axion may be relevant for the issue of the electroweak (EW) vacuum instability and therefore inflation. 

{\it The aim of this paper is to identify a simple and well-motivated model where the following signals of BSM physics can all be addressed and which adds to the SM only right-handed neutrinos and the extra fields needed to implement the axion idea:}
\begin{enumerate}
\item {\bf Small neutrino masses}. We adopt the perhaps simplest explanation: the {\it type-I see-saw mechanism} based on right-handed neutrinos. The addition of right-handed neutrinos also symmetrize the field content of the SM giving to each SM left-handed particle a right-handed counterpart.
\item {\bf Dark matter}. As a DM candidate we consider {\it the axion} \cite{Weinberg:1977ma}, a light spin-0 particle whose existence is implied by the spontaneous symmetry breaking of a U(1) symmetry, the Peccei-Quinn (PQ) symmetry \cite{Peccei:1977hh} that  explains why strong interactions do not violate CP. In particular, we consider the invisible axion model proposed by Kim, Shifman, Vainshtein and Zakharov (KSVZ)  \cite{Kim:1979if}, which has a simple structure and a small number of free parameters.
\item {\bf Baryon asymmetry}. In order to explain such asymmetry we make use of (thermal) {\it leptogenesis} \cite{fuk}, which is implemented with the same right-handed neutrinos that allow the light neutrinos to have masses.
\item {\bf Inflation and vacuum instability}.  As we stated before, the inflaton could be identified with the Higgs boson provided that the EW vacuum is stable\footnote{By adding non-renormalizable operators with independent coefficients one may enter the region of metastability \cite{Bezrukov:2014ipa}, we do not consider this possibility in the present paper.}, taking into account energies  up to the Planck scale. We therefore look for regions of the parameter space where the EW vacuum is stable, even for  central values of the SM observables.
\item {\bf Strong CP problem}. The solution we consider is the first and most famous one: {\it the PQ symmetry}, the same symmetry leading to the axion DM candidate above.
\end{enumerate}

It is important to note that the first two points represent a proof of BSM physics, while the others are indications, although very plausible ones. The spirit here is similar to the one of \cite{Asaka:2005pn}, focusing on problems 1, 2 and 3 and adding to the SM field content only right-handed neutrinos with masses below the EW scale.  In this case, indeed, the right-handed neutrinos can significantly contribute to dark matter \cite{Dodelson:1993je} (see also \cite{Kusenko:2009up} for a review and further references) and neutrino oscillations provide a mechanism to generate baryon asymmetry through a different version of leptogenesis \cite{Akhmedov:1998qx}. Moreover, it is possible to extend this framework to include the axion idea and to look for simultaneous solutions of problems 1, 2, 3, 4 and 5.  Here there is no claim that the simple model we study is the only one able to address all these issues. 

In the list above we did not include the gauge hierarchy and the cosmological constant problems because  they can both be addressed with anthropic arguments\footnote{The gauge hierarchy problem can of course be solved in a technically natural way (e.g. with SUSY, composite Higgs, etc) in models that explain some of the issues mentioned above \cite{Allahverdi:2006cx}; also,  SUSY large extra dimensions \cite{Aghababaie:2003wz}  offers a possible way to address the  cosmological constant problem; but this is done at the price of introducing many more fields than those of the model studied here and sometimes the necessity of an ultraviolet completion at much smaller energies. }  \cite{Weinberg:1987dv}. On the other hand, there seems to be no anthropic solution to the strong CP problem; thus technical naturalness appears to be the only possible way to explain the small value of the QCD  $\theta$ angle.

Let us summarize now the contents of the article. In section \ref{model} we define the model. In section \ref{constraints} we discuss the observational constraints  on its parameters.  The theoretical ingredients for the extrapolation up to $M_{\rm Pl}$ are provided in section \ref{RGE}. In section \ref{stab} we investigate whether the model can have a stable EW vacuum taking into account energies up to $M_{\rm Pl}$. One of the conditions for stability is that the Higgs quartic coupling remains always positive. There is, however, another condition to be fulfilled to ensure a stable vacuum. This section contains the central new results of this paper. Finally, in section \ref{conclusions} we provide our conclusions.

\section{The model} \label{model}
We consider the model with Lagrangian:
\be \mathcal{L} =   \mathcal{L} _{\rm gravity}+ \mathcal{L} _{\rm SM}+ \mathcal{L} _{N}+   \mathcal{L} _{\rm axion},\label{full-lagrangian}
 \ee
where repeated indices understand a summation.  The gauge group of the model is the SM one:  $$G_{\rm SM}= {\rm SU(3)_c\times SU(2)_{\it L}\times U(1)_{\it Y}}.$$ $\mathcal{L} _{\rm gravity}$ are the terms in the Lagrangian, which include the pure gravitational part and the possible non-minimal coupling between gravity and the other fields. In particular, the term proportional to $|H|^2 \mathcal{R}$, where $H$ is the Higgs doublet and $ \mathcal{R}$ is the Ricci scalar, plays an important role in Higgs inflation \cite{Bezrukov:2007ep}. $\mathcal{L} _{\rm SM}$ is the SM Lagrangian (minimally coupled to gravity).  $\mathcal{L} _{N}$ is the part of the Lagrangian that depends on the right-handed neutrinos $N_i$ (i=1,2,3):
\be \mathcal{L} _{N}= i\overline{N}_i \dsl N_i+ \left(\frac12 N_i M_{ij}N_j +  Y_{ij} L_iH N_j + {\rm h.  c.}\right),\ee
where $M_{ij}$ and $Y_{ij}$  are the elements of  the Majorana mass matrix $M$
and the neutrino Yukawa coupling matrix $Y$, respectively. Thanks to the complex Autonne-Takagi factorization,
we take $M$ real and diagonal without loss of generality: $$M=\mbox{diag}(M_1, M_2, M_3),$$ where the $M_i$  ($i=1,2,3$) are mass parameters, the Majorana masses of the three right-handed neutrinos.
 
Finally, $\mathcal{L} _{\rm axion}$ represents the additional terms in the Lagrangian due to the chosen axion model. As stated in the introduction, we consider the first invisible axion model (the KSVZ model\footnote{For a recent interesting work where another axion model and scalar generations of neutrino masses are considered, see \cite{Bertolini:2014aia}.}). The fields of this model that are not contained in the SM are the following.
\begin{itemize}
\item {\bf An extra Dirac fermion.} (In Weyl notation) it is a pair of two-component fermions $q_1$ and $q_2$ in the following representation of $G_{\rm SM}$ 
\be q_1 \sim (3,1)_0, \qquad q_2 \sim (\bar{3}, 1)_0.\ee
Namely they form a colored Dirac fermion   with no interactions with the gauge fields of $\rm SU(2)_{\it L}\times U(1)_{\it Y}$.
\item {\bf An extra complex scalar.} This scalar $A$ is charged under $\rm U(1)_{PQ}$ and neutral under $G_{\rm SM}$.
\end{itemize}
The Lagrangian of this axion model is 
\be \mathcal{L}_{\rm axion} = i\sum_{j=1}^2\overline{q}_j \Dsl \, q_j +|\partial_{\mu} A|^2  -(y\, q_2A q_1 +h.c.)-\Delta V(H,A)\nonumber \ee
and the classical potential of the full model is
\be  V(H,A)=  \lambda_H(|H|^2-v^2)^2+\Delta V(H,A), \ee
where
\be \Delta V(H,A) \equiv \lambda_A(|A|^2-  f_a^2)^2 + \lambda_{HA} (|H|^2-v^2)( |A|^2-f_a^2).\nonumber \ee
The parameters $v$, $f_a$ and $y$ can be taken real and positive without loss of generality. The PQ symmetry acts on $q_1$, $q_2$ and $A$ as follows
\be q_1\rightarrow e^{i\alpha/2}q_1, \quad q_2\rightarrow e^{i\alpha/2}q_2, \quad A\rightarrow e^{-i\alpha}A, \ee
which forbids  an explicit mass term  $M_q q_1 q_2  +h.c.\,$. The SM fields and the right-handed neutrinos are instead neutral under U(1)$_{\rm PQ}$. Moreover, there is the accidental symmetry
\be q_1\rightarrow  - q_1 ,\quad q_2\rightarrow  q_2,\quad A\rightarrow -A .
\ee 
This model has the advantage of being simple and having (in addition to the SM and type-I see-saw parameters) only three real parameters: $\lambda_{HA}$, $\lambda_A$ and $y$; it is the most general one given the field content and symmetries described above. In particular, notice that $\lambda_{HA}$ is the only tree-level  coupling between the axion and SM sectors.

The EW symmetry breaking is triggered by the vacuum expectation value (VEV) $v\simeq 174\,$GeV of the neutral component $H_0$ of the Higgs doublet. After that the neutrinos acquire a Dirac mass matrix
  \be m_D = v Y,\label{mDY}\ee 
  which can  be parameterized as 
\be m_D =\left(\begin{array}{ccc}\hspace{-0.1cm}m_{D1}\,, & \hspace{-0.2cm}m_{D2}\, ,  & \hspace{-0.2cm} m_{D3}\hspace{-0.1cm}
\end{array}\right), \label{Dirac-mass}\ee
where $m_{Di}$ ($i=1,2,3$) are column vectors.   Integrating out the heavy neutrinos $N_i$, one then obtains  the following light neutrino Majorana mass matrix  
\be m_\nu =  \frac{m_{D1} m_{D1}^T}{M_1} + \frac{m_{D2} m_{D2}^T}{M_2} + \frac{m_{D3} m_{D3}^T}{M_3} . \label{see-saw} \ee 
  By means of a unitary (Autonne-Takagi) redefinition of the left-handed SM neutrinos we can diagonalize $m_\nu$ to obtain the mass eigenvalues $m_1, m_2$ and $m_3$ (the left-handed neutrino Majorana masses). Calling $U_\nu$ the unitary matrix that implements such transformation, also known as the Pontecorvo-Maki-Nakagawa-Sakata (PMNS) matrix, that is $U^T_\nu m_\nu U_\nu= $ diag$(m_1, m_2, m_3)$, we can parameterize $U_\nu= V_\nu P_{12}$, where
\be 
\small V_\nu=\left(\begin{array}{ccc}
c_{12} c_{13} & s_{12} c_{13} & s_{13} e^{-i \delta}  \\ -s_{12}c_{23}-c_{12}s_{13}s_{23}e^{i\delta} & c_{12}c_{23}-s_{12}s_{13}s_{23}e^{i\delta}  & c_{13}s_{23} \\ 
s_{12}s_{23}-c_{12} s_{13} c_{23}e^{i\delta}  & -c_{12}s_{23}-s_{12}s_{13} c_{23} e^{i\delta}& c_{13}c_{23}
\end{array}\right),
\nonumber\ee
with $s_{ij}\equiv \sin(\theta_{ij})$, $c_{ij}\equiv \cos(\theta_{ij})$; $\theta_{ij}$ are the neutrino mixing angles and $P_{12}$ is a diagonal matrix that  contains two extra phases, in addition to the one, $\delta$, contained in $V_\nu$: 
\be P_{12}=\left(\begin{array}{ccc}
e^{i \beta_1} & 0  & 0  \\  0 & e^{i\beta_2} & 0\\ 
0  & 0 & 1
\end{array}\right).   \ee
Even in the most general case of three right-handed neutrinos, it is possible to express $Y$ in terms of low-energy observables, the heavy masses  $M_1$, $M_2$ and $M_3$ and extra parameters \cite{Casas:2001sr}: 
\be Y= \frac{U_\nu^* D_{\sqrt{m}}\, R \, D_{\sqrt{M}} }{v}, \label{Casas-Ibarra}\ee
where 
\bea D_{\sqrt{m}}&\equiv& \mbox{diag}(\sqrt{m_1},\sqrt{m_2},\sqrt{m_3}),\nonumber \\ D_{\sqrt{M}}&\equiv& \mbox{diag}(\sqrt{M_1},\sqrt{M_2}, \sqrt{M_3}) \nonumber\eea
and $R$ is a generic complex orthogonal matrix, which contains the extra parameters. This is useful for us because the observational constraints are not directly on $Y$, but they are rather on the low-energy quantities $m_i$, $U_\nu$ and on $M_i$ (see section \ref{constraints}).  One can show that the simpler and realistic case of  two right-handed neutrinos \cite{Ibarra:2005qi} below $M_{\rm Pl}$ can be recovered by setting $m_1=0$ and 
\be R = \left(\begin{array}{ccc}
0 & 0  & 1\\ \cos z & -\sin z & 0\\ 
\xi \sin z & \xi \cos z & 0
\end{array}\right),\nonumber\ee 
where $z$ is a complex parameter and $\xi=\pm 1$.

The PQ symmetry is broken both  spontaneously and by anomalies. The spontaneous symmetry breaking is induced by $f_a\equiv \langle A\rangle$, leading to the following Dirac mass of  $\{q_1,q_2\}$:  $$M_q = y f_a.$$ Moreover, $A$ contains a (classically) massless particle, the axion, which acquires a small mass thanks to the quantum breaking of the PQ symmetry, and  a massive particle with squared mass
\be M_A^2 = f_a^2\left(4\lambda_A +\mathcal{O}\left(\frac{v^2}{f_a^2}\right)\right).  \label{MA} \ee
As we will review below, the observational bounds imply that the corrections $\mathcal{O}\left(v^2/f_a^2\right)$ are very small and will be neglected in the following.

\section{Observational constraints} \label{constraints}

We now discuss the observational constraints, which we will take into account in the rest of the paper.

As far as the neutrino masses $m_i$ ($i=1,2,3$) are concerned, data from atmospheric and solar neutrinos tell us respectively   \cite{Gonzalez-Garcia:2014bfa}  (see also \cite{Tortola:2012te, Fogli:2012ua,GonzalezGarcia:2012sz} for previous determinations)
\bea  \Delta m^2_{21}&=& 7.50^{+0.19}_{-0.17}  \times 10^{-5} \,  {\rm eV}^2 , \nonumber\\   \Delta m^2_{3l} &=&2.457^{+0.047}_{-0.047} \times 10^{-3} \,  {\rm eV}^2, \nonumber \eea 
where $\Delta m_{ij}^2 \equiv m_i^2-m_j^2$ and $ \Delta m^2_{3l} \equiv  \Delta m^2_{31} $ for normal ordering and $ \Delta m^2_{3l} \equiv  \Delta m^2_{32} $ for inverted ordering.

As far as the mixing angles and phases of the PMNS matrix are concerned, the most recent central values and corresponding uncertainties can also be found  in \cite{Gonzalez-Garcia:2014bfa}: for any ordering of the neutrino masses the 3$\sigma$ ranges  are 
\bea 0.270 \leq s_{12}^2\leq 0.344 ,\quad  0.385 \leq s_{23}^2\leq 0.644, \nonumber\\ \quad 0.0188 \leq s_{13}^2\leq 0.0251, \eea
while $\delta$ spans the whole range from $0$ to $2\pi$ at 3$\sigma$ level (for example for normal ordering we have  $\delta /^0 = 306^{+39}_{-70}$, while, for inverted ordering, $\delta/^0=254^{+63}_{-62}$). Currently no significant constraints are known for $\beta_1$ and $\beta_2$.

We now turn to the requirements to have successful leptogenesis \cite{fuk}: neutrinos should be lighter than 0.15 eV and the lightest right-handed neutrino Majorana mass $M_l$ has to  fulfill  \cite{Davidson:2002qv}  
\be  M_l \gtrsim 1.7 	\times 10^{7}\, \mbox{GeV}.\label{leptobound} \ee
In order to be conservative we have reported the weakest bound, but depending on the assumptions one can have stronger conditions\footnote{For example if the initial abundance of right-handed neutrinos at $T\gg M_l$ is zero then the bound is  $M_l \gtrsim 2.4\times 10^{9}$ GeV  \cite{Davidson:2002qv}.}. Notice, however, that the mechanisms of \cite{Asaka:2005pn} and \cite{Akhmedov:1998qx} discussed in the introduction can evade these bounds and use right-handed neutrino masses below the EW scale; as we will see, it is less challenging to achieve vacuum stability in this case. In other models, if the Higgs field acquires a large VEV during inflation, \cite{Kusenko:2014lra} argued that the subsequent Higgs relaxation to the EW vacuum can generate the baryon asymmetry.

Regarding the axion sector,  in order to account for DM through the misalignment mechanism
 \cite{axionDM} (with an order one initial misalignment angle)  and to elude axion detection one obtains respectively an upper (see e.g.  \cite{Kawasaki:2013ae}) 
and lower bound (see e.g. \cite{Raffelt:1999tx}) on the order of magnitude of the scale of PQ symmetry breaking $f_a$:
\be 10^8 \, \mbox{GeV} \lesssim f_a \lesssim   10^{12} \, \mbox{GeV}. \label{bound-f}\ee
The upper bound is obtained by requiring that the axion field takes a value of order $f_a$ at early times, which is what we  expect, but is not necessarily the case; also the precise value of the lower bound is  model dependent. Therefore  (\ref{bound-f}) should not be taken as sharp bounds, but it certainly gives a plausible range of $f_a$. Another source of uncertainty is introduced if one instead considers light right-handed neutrinos  \cite{Dodelson:1993je,Kusenko:2009up}, which can then contribute to dark matter, as mentioned in the introduction; in this case, indeed, the upper bound becomes stronger as it is obtained by requiring the axion contribution not to exceed the observed dark matter abundance.  In any case, (\ref{bound-f})  ensures that $f_a\gg v$ and the terms $\mathcal{O}\left(v^2/f_a^2\right)$ in (\ref{MA}) can be neglected. Moreover, notice that bounds on $f_a$  can only constrain the ratio $M_A/\sqrt\lambda_A$ as it is clear from (\ref{MA}). When  $M_A\gg v$ and $M_q \gg v$ (which we assume) the EW constraints are fulfilled. 

In addition to contributing to dark matter, the axion also unavoidably manifest itself as dark radiation as it is also thermally produced  \cite{Masso, Graf:2010tv, Salvio:2013iaa}. This population of hot axions contributes to the effective number of relativistic species, but the size of this contribution is currently well within the observational bounds \cite{Salvio:2013iaa}.

 Finally, of course we also have  constraints on the SM parameters. After the discovery of the Higgs boson at the LHC \cite{Higgsexp1,Higgsexp2}  the last SM parameter, the Higgs mass, has been determined within small uncertainties, and there are no free SM parameters anymore. We take the values and uncertainties of the SM masses and couplings given in \cite{Buttazzo:2013uya} (see also the references therein). The determinations of  \cite{Buttazzo:2013uya} are not significantly affected by the presence of the extra heavy degrees of freedom.% as
 % $M_i\gg v$,
%  $M_A\gg v$ and $M_q \gg v$.

 \begin{figure}[t]
\includegraphics[scale=0.67]{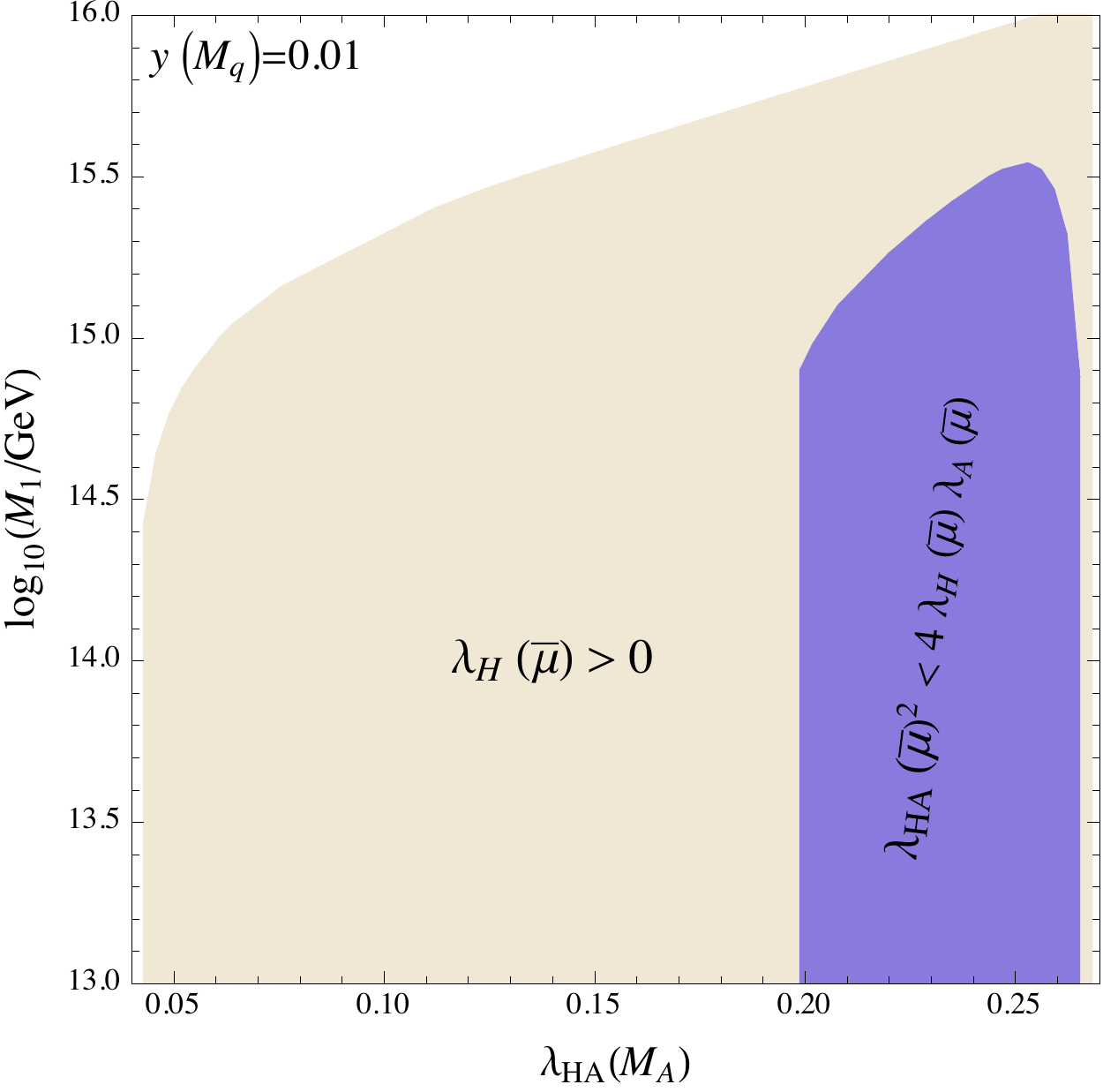} 
   \caption{{\small Phase diagram of the  model, showing the region with absolute stability up to the Planck scale. The region where condition II is fulfilled is inside the region with $\lambda_H (\bar \mu)>0$. We set the central values of the SM parameters at the EW scale and the low-energy neutrino parameters; however, we checked that  variations of $\Delta m_i^2$ and $\theta_{ij}$ (within 5$\sigma$ around their central values) and variations of  $\delta$,  $\beta_1$ and $\beta_2$  have a negligible effect on this plot. Moreover, we set the lightest neutrino mass $m_1=0$,  
    $M_2=10^{14}\mbox{GeV}$, $M_3 > M_{\rm Pl}$ and  $z=0$. Switching the sign of $\xi$ does not change the plot. The axion decay constant is set to  $f_a=10^{11}$GeV and $\lambda_A(M_A)=0.05$.}}
\label{PD}
\end{figure}

\section{RGE analysis and thresholds} \label{RGE}

Since we want to study the predictions of this model at energies much above the EW scale, up to the Planck scale, we need the complete set of renormalization group equations (RGEs).  We adopt the $\overline{\rm MS}$ renormalization scheme  to define the renormalized couplings and the corresponding RGEs. Moreover, for a generic renormalized coupling $g$ we write the RGEs as
\be \frac{dg}{d\tau}= \beta_{g},\ee
where $d/d\tau\equiv \bar{\mu}^2\, d/d\bar{\mu}^2$ and $\bar{\mu}$ is the $\overline{\rm MS}$ renormalization energy scale. The $\beta$-functions  $\beta_{g}$ can also be expanded in loops as 
\be  \beta_{g} =  \frac{\beta_{g}^{(1)}}{(4\pi)^2}+ \frac{\beta_{g}^{(2)}}{(4\pi)^4}+ ... \, ,\ee 
where $ \beta_{g}^{(n)}/(4\pi)^{2n}$   is the $n$-loop contribution. 

Let us start from energies much above $M_A$, $M_q$ and $M_{ij}$. In this case the 1-loop RGEs   are (see \cite{Pirogov:1998tj,EliasMiro:2011aa,EliasMiro:2012ay,Salvio:2014soa} for previous determinations of some terms in these RGEs)
%
%\small
\bea  \beta_{g_1^2}^{(1)}& =&    \frac{41g_1^4}{10}, \qquad   \beta_{g_2^2}^{(1)} =- \frac{19g_2^4}{6},\qquad\beta_{g_3^2}^{(1)}  = -\frac{19 g_3^4}{3},\nonumber\\   \beta_{y_t^2}^{(1)}  & =& y_t^2\left(\frac92 y_t^2-8g_3^2-\frac{9g_2^2}{4}-\frac{17g_1^2}{20} + {\rm Tr}(Y^\dagger Y )\right),\nonumber\\ 
  \beta_{\lambda_H}^{(1)} & =&\left(12\lambda_H+6y_t^2-\frac{9g_1^2}{10}-\frac{9g_2^2}{2}+2\, {\rm Tr}(Y^\dagger Y)\right)\lambda_H \nonumber\\ &&\hspace{-0.7cm}-\, 3y_t^4 +\frac{9 g_2^4}{16}+\frac{27 g_1^4}{400}+\frac{9 g_2^2 g_1^2}{40}+\frac{\lambda_{HA}^2}{2} - {\rm Tr}((Y^\dagger Y)^2), \nonumber\\ 
 \beta_{\lambda_{HA}}^{(1)} & =& \left(3y_t^2-\frac{9g_1^2}{20}-\frac{9g_2^2}{4}+6\lambda_H \right) \lambda_{HA}\nonumber \\ && +\left(4\lambda_A +\, {\rm Tr}(Y^\dagger Y ) + 3y^2 \right) \lambda_{HA}+2 \lambda_{HA}^2, \nonumber\\ 
 \beta_{\lambda_A}^{(1)} & =& \lambda_{HA}^2+10\lambda_A^2+6y^2 \lambda_A- 3 y^4,\nonumber\\ 
   \beta_{Y}^{(1)} & =&Y  \left[\frac32 y_t^2-\frac{9}{40} g_1^2-\frac98 g_2^2+\frac34 Y^\dagger Y+\frac12 {\rm Tr}(Y^\dagger Y )\right],\nonumber \\ 
 \beta_{y^2}^{(1)} & =&y^2(4y^2-8 g_3^2),\nonumber
\eea
where  $g_3$,  $g_2$ and  $g_1=\sqrt{5/3}g_Y$ are the gauge couplings of ${\rm SU(3)}_c $, ${\rm SU(2)_{\it L}}$ and   ${\rm U(1)_{\it Y}}$ respectively and $y_t$ is the top Yukawa coupling.  The explicit form of the complete set of the RGEs above was not explicitly presented before, but the RGEs for a generic quantum field theory (without gravity) were computed up to 2-loop order in \cite{MV} (see also \cite{Lyonnet:2013dna} for a computer implementation of them).

Next, we consider what happens in going from energies above $M_A$  to energies below $M_A$: as discussed in \cite{RandjbarDaemi:2006gf,EliasMiro:2012ay} one has to take into account a scalar threshold effect: in the low energy effective field theory below $M_A$  one has the effective Higgs quartic coupling 
\be \lambda= \lambda_H-\frac{\lambda_{HA}^2}{4 \lambda_A}. \label{lambda_eff}\ee
This is the result of integrating out the massive scalar  degree of freedom at  tree-level. The reason why this shift  occurs is  because setting the heavy scalar to zero is not a consistent truncation, namely it is not consistent with the equations of motion. In practice one should do the following: below $M_A$ the RGEs are  the ones  given above with $\beta_{\lambda_{HA}}$ and $\beta_{\lambda_{A}}$ removed and  $\lambda_H$ replaced by $\lambda$. Above $M_A$ one should include  $\beta_{\lambda_{HA}}$ and $\beta_{\lambda_{A}}$ and find $\lambda_H$ using the full RGEs and the boundary condition in (\ref{lambda_eff}) at $\bar \mu= M_A$.

As far as the new fermions are concerned, 
following  \cite{Casas:1999cd} we adopt the approximation in which the new Yukawa couplings run only above the corresponding mass thresholds; this is implemented technically by substituting $Y_{ij} \rightarrow Y_{ij} \theta(\bar{\mu}-M_j)$ and $y	\rightarrow y \theta(\bar{\mu} - M_q)$ on the right-hand side of the RGEs. The situation is different from the scalar one, as setting the fermion fields to zero below their mass threshold is consistent. 

Finally notice that, the SM parameters can run in an energy range bigger than the one  of $Y$, $\lambda_A$, $\lambda_{HA}$ and $y$. Therefore, we include for them the 2-loop RGE contribution; we do not, however, show explicitly the 2-loop part because of its complexity.

\section{Stability analysis} \label{stab}

Since we use the 1-loop RGEs of the non-SM parameters, we approximate the  Coleman-Weinberg \cite{CW}  effective potential  of the model with its  RG-improved tree-level potential: we substitute the bare couplings in the classical potential with the corresponding running ones. 

The conditions that ensure the absolute stability  of the vacuum $\langle H_0\rangle = v$ and $\langle A\rangle =f_a$   have been studied in 
\cite{EliasMiro:2012ay}: they are
\begin{description}
\item[I.] $\lambda_H(\bar{\mu}) >0$ and $\lambda_A(\bar{\mu}) >0$
\item[II.]  $\Lambda_c\equiv 4\lambda_H(\bar{\mu}) \lambda_A(\bar{\mu}) -\lambda_{HA}^2(\bar{\mu}) >0$
\end{description}
Notice that once $\lambda_H>0$ and $\Lambda_c>0$ are fulfilled then $\lambda_A>0$ is fulfilled too. The fact that the $\overline{\rm MS}$ couplings are gauge invariant, as proved in \cite{wil,Buttazzo:2013uya}, guarantees that our results will not be affected by any gauge dependence.

\begin{figure}[t]
\includegraphics[scale=0.67]{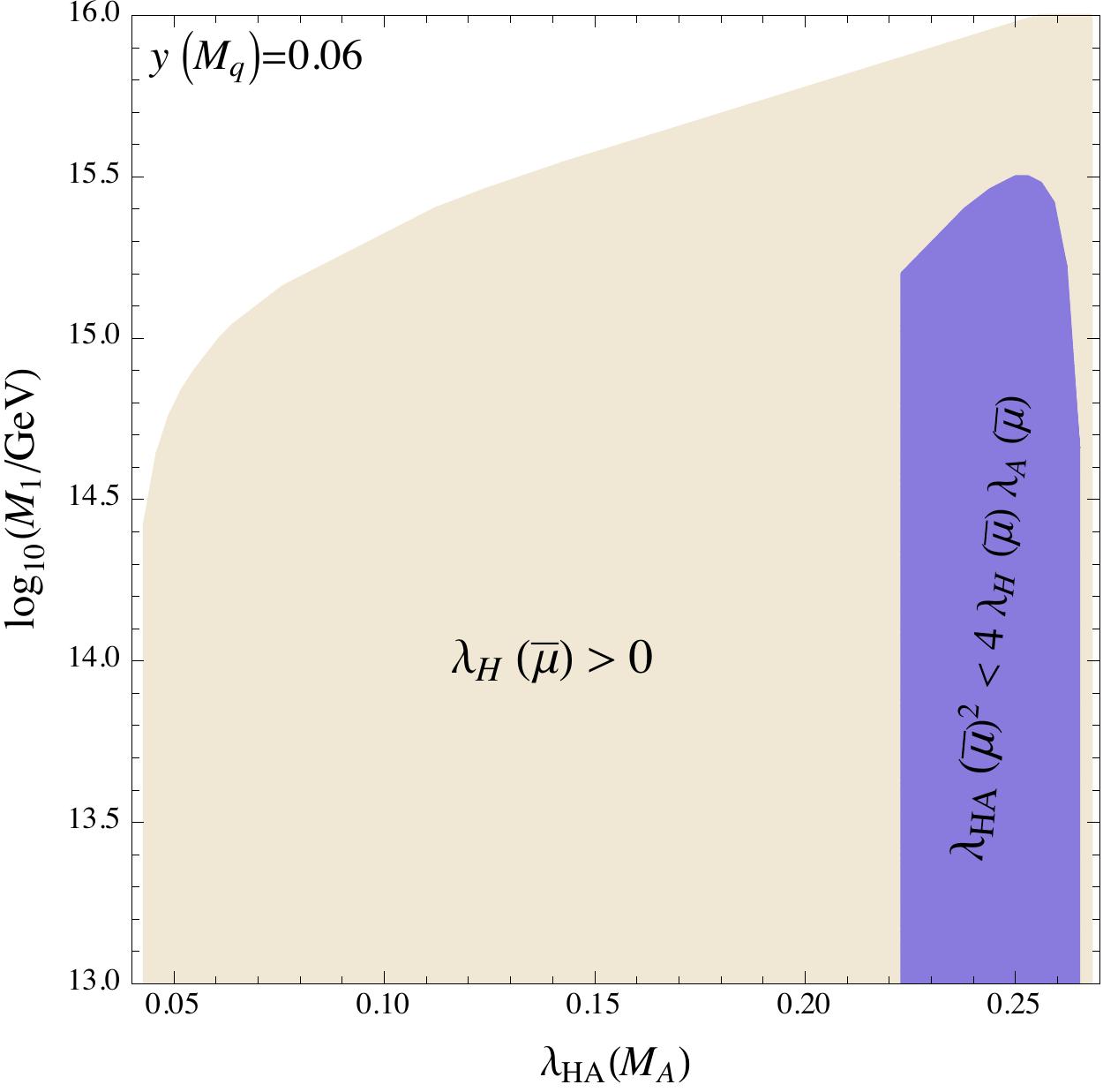} 
   \caption{{\small The same as in fig. \ref{PD}, but with a different value of $y$.}}
\label{PD2}
\end{figure}
\begin{figure}[t]
\includegraphics[scale=0.68]{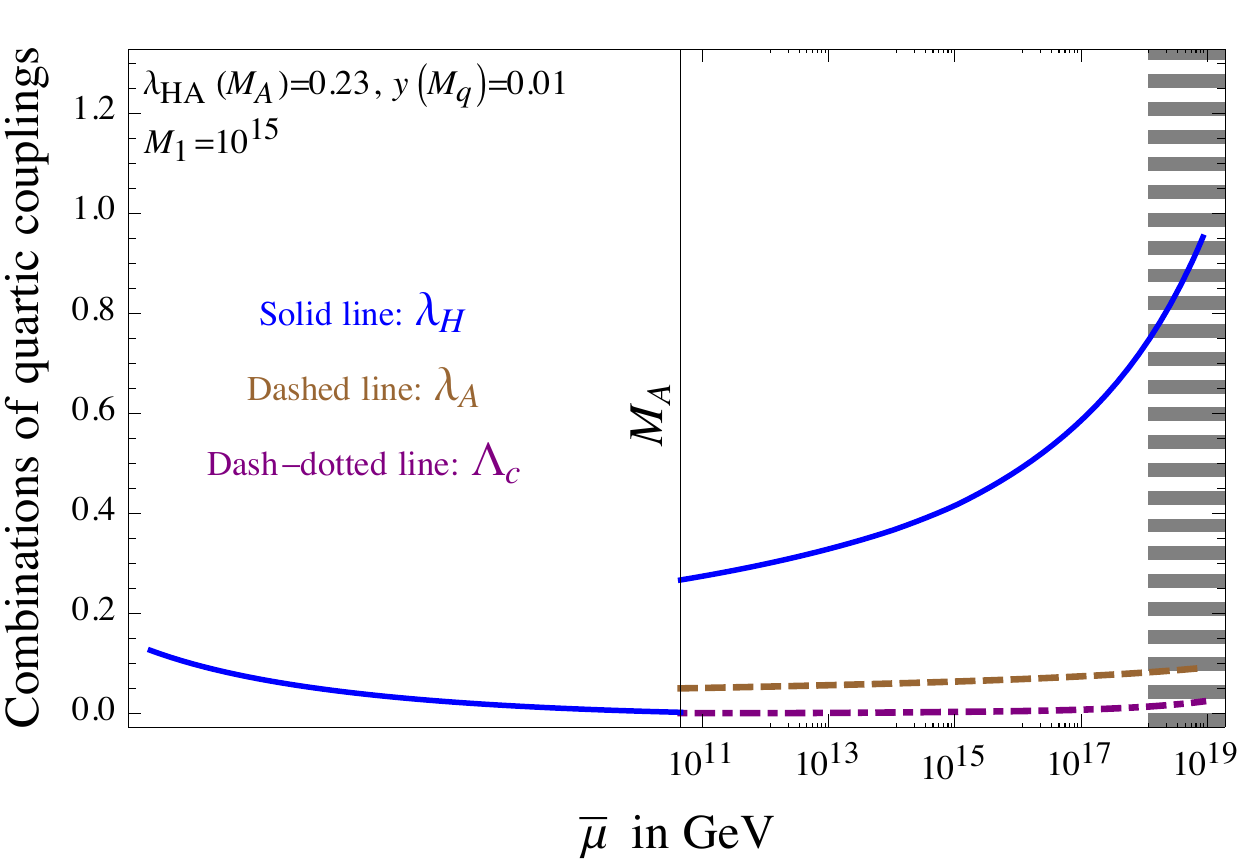} 
   \caption{{\small RG evolution of the quartic couplings $\lambda_H$, $\lambda_A$ and the combination of quartics $\Lambda_c$ defined in condition II for stability. The vertical solid line indicates the position of the scalar threshold, $M_A$. The stripes on the right indicate the region presumably dominated by Planck physics.  The values of the parameters are the same used in figure \ref{PD}.}}
\label{evolution1}
\end{figure}

 \begin{figure}[t]
\includegraphics[scale=0.68]{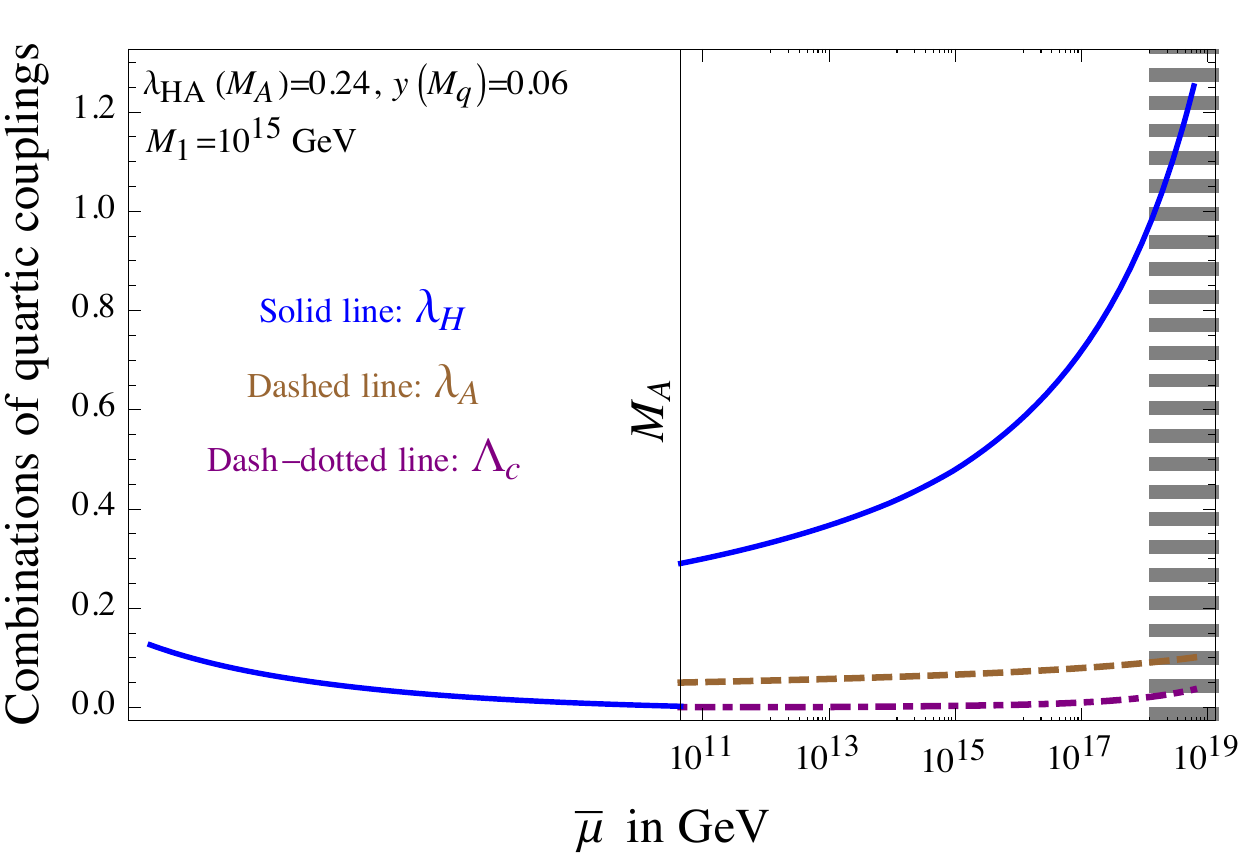} 
   \caption{{\small The same plot as in fig. \ref{evolution1}, but with the values of the parameters used in figure \ref{PD2}.}}
\label{evolution2}
\end{figure}

The first condition $\lambda_H>0$, at the level of approximation we are using, may lead to the possibility of Higgs inflation \cite{Bezrukov:2008ej, Bezrukov:2012sa,Salvio:2013rja}. Therefore having absolute stability may also allow us  to identify the inflaton with the Higgs field. However, one should keep in mind that perturbative unitarity\footnote{This unitarity problem can be solved by adding an extra real scalar field \cite{Giudice:2010ka,EliasMiro:2012ay}. The extension of the present analysis to include such scalar is beyond the scope of this paper.} is violated above some high energy scale \cite{crit,Burgess:2010zq}. Once the background fields are taken into account, however,  the authors of \cite{Bezrukov:2010jz} find that such energy is parametrically higher than all relevant scales during the history of the Universe. Nevertheless some extra assumptions on the underlying ultraviolet completion are necessary  \cite{Burgess:2010zq,Bezrukov:2010jz, Bezrukov:2012sa}. 

The question of the stability of the EW vacuum has been addressed previously in other economic extensions of the SM. The SM extended only by adding a single right-handed neutrino or three right-handed neutrinos with degenerate masses was studied in \cite{Casas:1999cd, EliasMiro:2011aa}. Extensions with a  singlet scalar were considered in \cite{Gonderinger:2009jp,EliasMiro:2012ay,Kadastik:2011aa} and others with one right-handed neutrino and an extra real scalar were studied in \cite{Chen:2012faa}. However, we do not know of any previous work that accounted for {\it all} problems listed in the introduction\footnote{After posting this article on the arXiv our attention was drawn to the interesting Ref. \cite{Dias:2014osa}. The authors discuss a model very similar to ours and anticipate that all those problems (with the exception of the origin of inflation) can be solved: in that work the PQ symmetry is an extension of the SM lepton number. This allows to relate the scales $f_a$ and $M_i$ \cite{Langacker:1986rj}. However, an explicit analysis was not presented in \cite{Dias:2014osa}. As we will see now, such an analysis here leads to regions where the simultaneous solutions occur and others where they do not.}.

In fig.  \ref{PD} and \ref{PD2} we show regions of the parameter space where the stability conditions are fulfilled for all values of $\bar \mu$ up to $M_{\rm Pl}$ and others where they are not. The values of the parameters used in that plot can also explain   neutrino masses, dark matter, baryon asymmetry and the strong CP problem (through the mechanisms discussed in the introduction), fulfilling all bounds of section \ref{constraints}. Moreover, the regions where $\lambda_H> 0$ all the way up to $M_{\rm Pl}$ correspond to the possibility of Higgs inflation. In fig. \ref{PD2} we see that increasing $y(M_q)$ shrinks the region where   condition II for stability is fulfilled: this is because $y$ contributes positively (negatively) to the running of $\lambda_{HA}$ ($\lambda_A$), which then increases (decreases) and this  makes it more difficult to satisfy that condition. We also observed that changing the value of $\lambda_A(M_A)$ and $f_a$ changes the location of that region, so that the size of the parameter space that is compatible with absolute stability is larger. Notice that figs  \ref{PD} and \ref{PD2} also indicate that lighter right-handed neutrino masses favor the stability conditions. This can be qualitatively understood: smaller $M_i$ generically correspond to smaller $Y_{ij}$, Eq. (\ref{Casas-Ibarra}), and to a reduced destabilizing effect in conditions I and II because of the way $Y$ appears in $ \beta_{\lambda_H}^{(1)}$ and $  \beta_{\lambda_{HA}}^{(1)}$.

In figs. \ref{evolution1} and \ref{evolution2} we show the evolution of the quartic coupling combinations relevant for the stability analysis as a function of the renormalization scale.  The parameters are chosen in a way compatible with the regions of, respectively, figs. \ref{PD} and \ref{PD2}, where all stability conditions are fulfilled. There are no Landau poles below the Planck scale and the couplings remain perturbative when the stability conditions are fulfilled. The region with stripes on the right corresponds to the regime where Planck physics is expected to be dominant; the behavior of the curves there is thus presumably unreliable.

At the same time, it is important to notice that there are also regions of the parameter space,  where the results on the stability analysis  obtained in the SM are not significantly changed by the addition of $N_i$, $q_j$ and $A$. In the limit $\lambda_{HA}\rightarrow 0$ the axion sector is decoupled from the rest, and, if the neutrino Yukawa couplings are small enough, one recovers the SM results at a very good level of accuracy.

\section{Conclusions}\label{conclusions}

In this paper we have found regions of the parameter space of a simple but well-motivated model that  can account for all experimentally confirmed signals of physics beyond the SM: neutrino oscillations (through the addition of three right-handed neutrinos), dark matter (due to the axion), baryon asymmetry (generated by thermal leptogenesis), inflation (which could be driven by the Higgs field since the EW vacuum can be an absolute minimum for energies up to the Planck scale) and the strong CP problem that is automatically solved by the PQ symmetry leading to the axion. 

This model is an extension of the SM, which only adds to the SM three right-handed neutrinos as well the scalar field and extra colored fermion of the simple invisible axion model proposed by KSVZ.

We have found that there are values of the parameters such that the important features listed above are all present together with perturbativity (always up to the Planck scale).

An important extension for the present work may be the inclusion of quantum gravity, which has been completely neglected here. Some steps in this direction have been taken in  \cite{Salvio:2014soa}. But the role of gravitational quantum effects in the stability issue of the SM is still unclear.

\section*{Acknowledgments} 
\noindent I thank J. Alberto Casas, Michele Frigerio, Thomas Hambye, Michele Maltoni, Mikhail Shaposhnikov  and 	C\'edric Weiland for very useful  discussions.   This work has been supported by the Spanish Ministry of Economy and Competitiveness under grant FPA2012-32828,  Consolider-CPAN (CSD2007-00042), the grant  SEV-2012-0249 of the ``Centro de Excelencia Severo Ochoa'' Programme and the grant  HEPHACOS-S2009/ESP1473 from the C.A. de Madrid.

%\label{ciiiiiii}



%% References


%\bibliographystyle{elsarticle-num}



%\vspace{1cm}

%\noindent{\bf References}

\end{document}